\documentclass[twocolumn,english,superscriptaddress,citeautoscript,showpacs,preprintnumbers,amsmath,amssymb,prl,floatfix,footinbib]{revtex4}
\usepackage[scaled=2]{helvet}
\usepackage[T1]{fontenc}
\usepackage[latin9]{inputenc}
\usepackage{array}
\usepackage{textcomp}
\usepackage{multirow}
\usepackage{amsmath}
\usepackage{amssymb}
\usepackage{graphicx}

\makeatletter

\providecommand{\tabularnewline}{\\}

\@ifundefined{textcolor}{}
{%
 \definecolor{BLACK}{gray}{0}
 \definecolor{WHITE}{gray}{1}
 \definecolor{RED}{rgb}{1,0,0}
 \definecolor{GREEN}{rgb}{0,1,0}
 \definecolor{BLUE}{rgb}{0,0,1}
 \definecolor{CYAN}{cmyk}{1,0,0,0}
 \definecolor{MAGENTA}{cmyk}{0,1,0,0}
 \definecolor{YELLOW}{cmyk}{0,0,1,0}
}

\usepackage[scaled=2]{helvet}\usepackage{amstext}

\makeatletter
\@ifundefined{textcolor}{}{\definecolor{BLACK}{gray}{0}\definecolor{WHITE}{gray}{1}\definecolor{RED}{rgb}{1,0,0}\definecolor{GREEN}{rgb}{0,1,0}\definecolor{BLUE}{rgb}{0,0,1}\definecolor{CYAN}{cmyk}{1,0,0,0}\definecolor{MAGENTA}{cmyk}{0,1,0,0}\definecolor{YELLOW}{cmyk}{0,0,1,0}}

\usepackage{dcolumn}
\usepackage{bm}
\usepackage{times}
\usepackage{hyperref}
\hypersetup{colorlinks=true,linkcolor=red,citecolor=blue}
\makeatother

\usepackage{babel}

\makeatother

\usepackage{babel}
\begin{document}

\title{Magnetization distribution and orbital moment in the non-Superconducting
Chalcogenide Compound K$_{0.8}$Fe$_{1.6}$Se$_{2}$}

\author{S. Nandi}

\email{s.nandi@fz-juelich.de}

\selectlanguage{english}%

\affiliation{Jülich Centre for Neutron Science JCNS and Peter Grünberg Institut
PGI, JARA-FIT, Forschungszentrum Jülich GmbH, D-52425 Jülich, Germany}

\affiliation{Jülich Centre for Neutron Science JCNS, Forschungszentrum Jülich
GmbH, Outstation at MLZ, Lichtenbergstraße 1, D-85747 Garching, Germany}

\author{Y. Xiao}

\affiliation{Jülich Centre for Neutron Science JCNS and Peter Grünberg Institut
PGI, JARA-FIT, Forschungszentrum Jülich GmbH, D-52425 Jülich, Germany}

\author{Y. Su}

\affiliation{Jülich Centre for Neutron Science JCNS, Forschungszentrum Jülich
GmbH, Outstation at MLZ, Lichtenbergstraße 1, D-85747 Garching, Germany}

\author{L. C. Chapon}

\affiliation{Institut Laue-Langevin, BP 156, 38042 Grenoble Cedex 9, France}

\author{T. Chatterji}

\affiliation{Institut Laue-Langevin, BP 156, 38042 Grenoble Cedex 9, France}

\author{W. T. Jin}

\affiliation{Jülich Centre for Neutron Science JCNS and Peter Grünberg Institut
PGI, JARA-FIT, Forschungszentrum Jülich GmbH, D-52425 Jülich, Germany}

\affiliation{Jülich Centre for Neutron Science JCNS, Forschungszentrum Jülich
GmbH, Outstation at MLZ, Lichtenbergstraße 1, D-85747 Garching, Germany}

\author{S. Price}

\affiliation{Jülich Centre for Neutron Science JCNS and Peter Grünberg Institut
PGI, JARA-FIT, Forschungszentrum Jülich GmbH, D-52425 Jülich, Germany}

\author{T. Wolf}

\affiliation{Institut für Festkörperphysik, Karlsruhe Institute of Technology,
D-76021 Karlsruhe, Germany}

\author{P. J. Brown}

\affiliation{12 Little St. Mary's Lane, Cambridge, CB2\,1RR, UK}

\affiliation{Institut Laue-Langevin, BP 156, 38042 Grenoble Cedex 9, France}

\author{Th. Brückel}

\affiliation{Jülich Centre for Neutron Science JCNS and Peter Grünberg Institut
PGI, JARA-FIT, Forschungszentrum Jülich GmbH, D-52425 Jülich, Germany}

\affiliation{Jülich Centre for Neutron Science JCNS, Forschungszentrum Jülich
GmbH, Outstation at MLZ, Lichtenbergstraße 1, D-85747 Garching, Germany}
\begin{abstract}
We have used polarized and unpolarized neutron diffraction to determine
the spatial distribution of the magnetization density induced by a
magnetic field of 9 T in the tetragonal phase of K$_{0.8}$Fe$_{1.6}$Se$_{2}$.
The maximum entropy reconstruction shows clearly that most of the
magnetization is confined to the region around the iron atoms whereas
there is no significant magnetization associated with either Se or
K atoms. The distribution of magnetization around the Fe atom is slightly
nonspherical with a shape which is extended along the $\left\langle 0\,0\,1\right\rangle $
direction in the projection. Multipolar refinement results show that
the electrons which give rise to the paramagnetic susceptibility are
confined to the Fe atoms and their distribution suggests that they
occupy 3\emph{d} \emph{t}$_{2g}$-type orbitals with around 66\% in
those of \emph{xz/yz} symmetry. Detail modeling of the magnetic form
factor indicates the presence of an orbital moment to the total paramagnetic
moment of Fe$^{2+}$.
\end{abstract}
\maketitle

\section{Introduction}

The discovery of iron-based superconductors \cite{kamihara_08} a
few years ago has stimulated tremendous research interests worldwide
in unconventional high-\emph{T}$_{\textup{C}}$ superconductivity.
The new excitement in this field has been generated very recently
due to the discovery of the new superconducting compound K$_{x}$Fe$_{2-y}$Se$_{2}$
with superconducting transition temperature, \emph{T}$_{\textup{C}}$,
above 30 K \cite{Guo_10}. Isostructural \emph{A}$_{x}$Fe$_{2-y}$Se$_{2}$
compounds (\emph{A} = Rb, Cs, and Tl \cite{Wang_11,Maziopa_11,Fang_11})
with similar \emph{T}$_{\textup{C}}$ have been found soon after.
One of the fascinating properties of the K$_{x}$Fe$_{2-y}$Se$_{2}$
superconductors, in contrast to the previously discovered pnictide
or chalcogenide superconductors, is the absence of the hole Fermi
surface at the Brillouin zone center or the presence of electronic
Fermi surface at the zone center \cite{Zhang,Wang_11,Qian_11,Mou_11}.
This poses a serious challenge to the well accepted theories of the
prevailing \emph{s}$^{\pm}$ pairing symmetry driven by the interband
scattering as suggested in many weak coupling theories \cite{Mazin_09}.
Another unusual feature of the K$_{x}$Fe$_{2-y}$Se$_{2}$ superconductors
is the presence of an antiferromagnetic order with a large moment
($\sim$\,3.3\,$\mu_{B}$) and very high transition temperature
($\sim$\,600 K) \cite{Bao_11} which is in contrast to the parent
compound of the pnictide superconductors where Fe moments order antiferromagnetically
at considerably lower temperature ($\sim$150 K) and with the small
ordered magnetic moment ($\sim$ 0.5-0.8 $\mu_{B}$) \cite{cruz_08,Johnston}.
Initially it was suggested that the superconductivity and antiferromagnetism
coexist and compete within the same phase of the K$_{x}$Fe$_{2-y}$Se$_{2}$
\cite{Bao_11}. However, subsequent detailed investigations concluded
a phase separation between the vacancy ordered antiferromagnetic phase
and a superconducting phase. Based on the observation of the $\sqrt{5}\times\sqrt{5}$
superlattice in the vacancy ordered antiferromagnetic phase, optimal
composition of \emph{A}$_{2}$Fe$_{4}$Se$_{5}$ has been suggested
for the parent phase \cite{Bao_11,Ye_11}. The nature of the superconducting
phase is still not settled yet. Both a vacancy free phase with composition
KFe$_{2}$Se$_{2}$ \cite{Ricci,Yuan_11,Li_12,Ding_13} and a phase
\emph{A}$_{2}$Fe$_{7}$Se$_{8}$ \cite{Ding_13} with Fe vacancies
have been found and was assigned to the superconducting phase.

Orbital composition of Fermi surface is very important regarding the
pairing mechanism and pairing strength for the Fe-based superconductors.
It has been shown theoretically that the strong interorbital interaction
is very efficient to achieve superconductivity due to magnetic fluctuations
in iron pnictides owing to the distinct orbital character of the Fermi
surface \cite{Zhang_09}. Besides the superconducting properties,
physical properties of the Fe-based superconductors are also strongly
dependent on the orbital character and occupancies of the Fe \emph{d}-orbitals.
Indeed, for the superconducting K$_{x}$Fe$_{2-y}$Se$_{2}$ ($x\thicksim0.76$,
$y\thicksim0.22$) compounds, a crossover from a low temperature metallic
state to an orbital selective Mott phase at high temperature has been
observed using Angle-Resolved Photoemission Spectroscopy (ARPES) measurements
\cite{Yi_13}. ARPES measurements clearly show that the spectral weight
of the Fe \emph{d}$_{xy}$ orbital near the Fermi surface is diminished
at high temperature while the \emph{d}$_{xz}/$\emph{d}$_{yz}$ orbitals
remain metallic. In order to obtain direct information about the electronic
states near the Fermi surface we have undertaken magnetization distribution
study of the non-superconducting compound K$_{0.8}$Fe$_{1.6}$Se$_{2}$
using polarized neutron diffraction.

\section{Experimental Details}

A good quality single crystal with approximate mass of 300 mg was
grown by the Bridgman method \cite{Landsgesell_12}. The structural
parameters were determined from unpolarized neutron diffraction measurement
using the four-circle diffractometer D9 equipped with a Cu (2 2 0)
monochromator to produce a monochromatic neutron beam of 0.838\,Å.
The flipping ratios were measured using the polarized neutron diffractometer
D3 with neutron wavelength of 0.825\,Å obtained with a Heusler alloy
monochromator with polarization of the incident neutron beam \emph{P}$_{0}$\,=\,0.907(5).
Both these instruments are installed on the hot neutron source of
the high-flux reactor of the Institute Laue-Langevin in Grenoble.
The sample was held at constant temperature in a closed-cycle refrigerator
on D9 whereas on D3 it was oriented with a $\left\langle 1\,1\,0\right\rangle $
axis parallel to the vertical field direction of a 9 T cryomagnet.
The crystallographic notations used here and in the rest of the paper
are according to the high temperature tetragonal phase with the \emph{I}4/\emph{mmm}
symmetry. The flipping ratios from the K$_{0.8}$Fe$_{1.6}$Se$_{2}$
crystal were measured in the paramagnetic tetragonal phase at \emph{T}\,=\,600
K. In a flipping ratio measurement, one measures the ratio $R=\frac{I^{+}}{I^{-}}$,
where I$^{+}$ and I$^{-}$ are the scattered neutron intensities
with neutron polarizations parallel and antiparallel to the applied
magnetic field directions, respectively. Because the induced moment
is small in K$_{0.8}$Fe$_{1.6}$Se$_{2}$, in the limit ($\gamma r_{0}/2\mu_{B})F_{M}(\mathbf{Q})/F_{N}(\mathbf{Q})$$\ll$1,
the flipping ratio, \emph{R}, can be expressed as \cite{Lester_11}
\begin{equation}
R\approx1-\frac{2\gamma r_{0}}{\mu_{B}}\frac{F_{M}(\mathbf{Q})}{F_{N}(\mathbf{Q})}\label{eq:R}
\end{equation}
where, $\gamma r_{0}=5.36\times10^{-15}$\,m and $\mu_{B}$ is the
Bohr magneton. \emph{F}$_{M}(\mathbf{Q})$ and \emph{F}$_{N}(\mathbf{Q})$
are the nuclear and the magnetic structure factors at the reciprocal
lattice vector \textbf{Q}. Since, \emph{F}$_{N}(\mathbf{Q})$ and
$R$ are known from the unpolarized and polarized neutron diffractions,
respectively, \emph{F}$_{M}(\mathbf{Q})$ can be calculated.

\section{Experimental Results}

\subsection{Macroscopic Characterizations}

\begin{figure}
\centering{}\includegraphics[clip,width=0.45\textwidth]{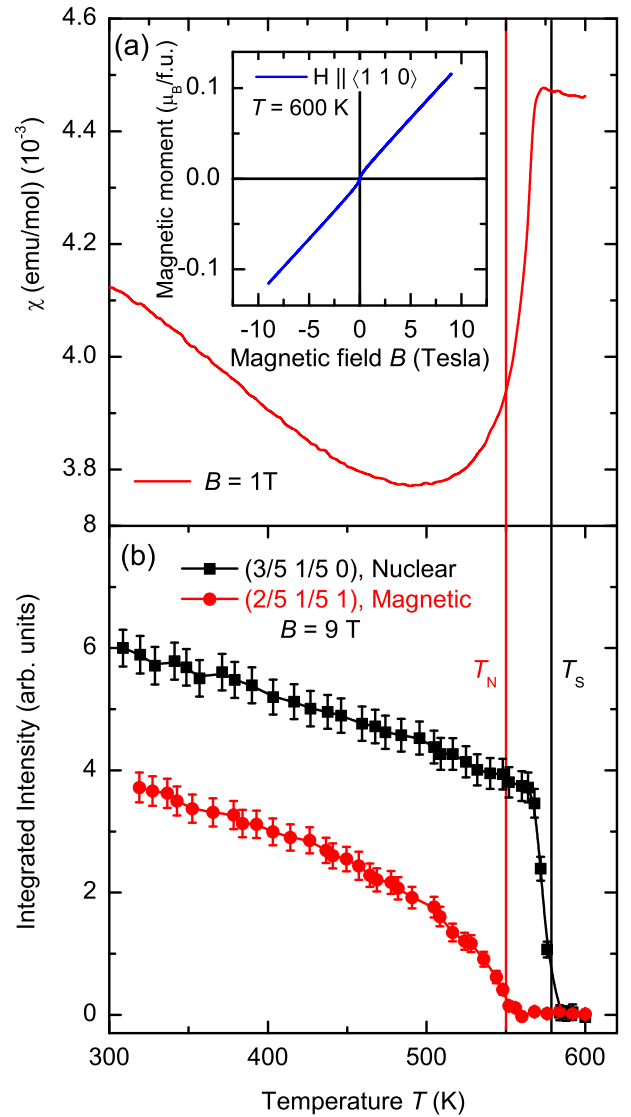}\\
 \caption{\label{fig1}(a) (Color online) Temperature dependence of the magnetic
susceptibility measured on heating of the sample in a field of 1\,T.
Inset shows \emph{M-H} curve for the same sample at \emph{T} = 600
K, above the magnetic ordering transition temperature of Fe. (b) Temperature
dependence of the nuclear ($\frac{3}{5}$ $\frac{1}{5}$ 0) and magnetic
($\frac{2}{5}$ $\frac{1}{5}$ 1) reflections, signaling the onset
of structural and magnetic phase transitions at \emph{T}\,=\,580
and 553 K, respectively.}
\end{figure}

\begin{figure}
\centering{}\includegraphics[clip,width=0.5\textwidth]{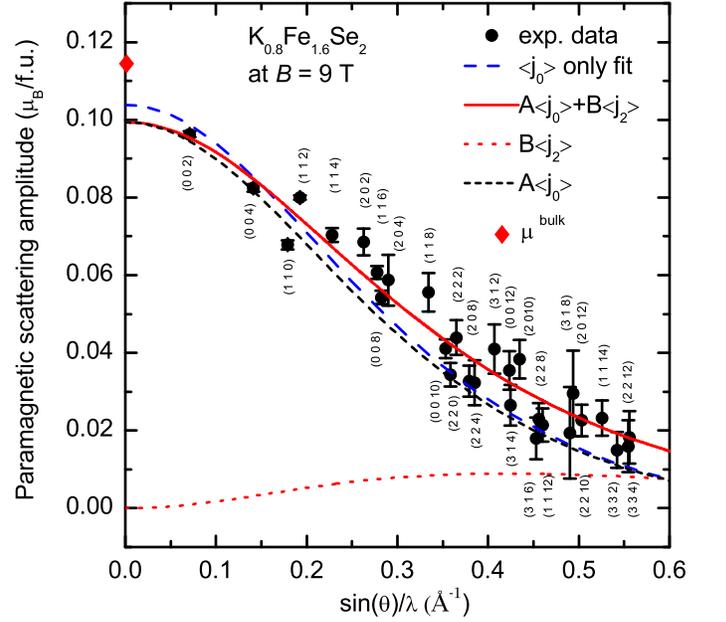}\\
 \caption{\label{fig2} (Color online) Paramagnetic scattering amplitudes of
Fe at \emph{T}\,=\,600 K. The large-dashed curve (blue) shows fitting
using the $\left\langle j_{0}\right\rangle $ form factor for Fe$^{2+}$,
Ref. \onlinecite{Freeman_61}. The solid (red) curve shows fitting
with $\left\langle j_{0}\right\rangle $ and $\left\langle j_{2}\right\rangle $
form factors with individual contributions are indicated by short-dashed
(black) and dotted (red) lines, respectively \cite{magnetic_form_factors}.
A and B are fitting parameters.}
\end{figure}

Figure \ref{fig1} (a) shows magnetic susceptibility of a single crystal
of K$_{0.8}$Fe$_{1.6}$Se$_{2}$ measured using a Quantum Design
vibrating sample magnetometer (VSM). For the magnetization measurements
a sample from the same batch as the diffraction measurements was used.
Magnetic susceptibility shows a clear kink at 572$\pm2$\,K, indicating
a phase transition from the high temperature vacancy disordered phase
to the low temperature vacancy ordered phase. Inset to the Fig.\,\ref{fig1}
(a) shows \emph{M-H }measured at \emph{T} = 600 K. The linear behavior
of the \emph{M-H} curve confirms the paramagnetic nature of the sample
at this temperature. The magnetization induced by a field of 9 T applied
along the $\left\langle 1\,1\,0\right\rangle $ direction at 600 K
was measured as 0.101(2)\,$\mu_{B}$/f.u. %
\footnote{In this paper, we have used one formula unit as one crystallographic
unit cell. Therefore, in one formula unit, on the average there are
3.6 Fe atoms considering the Fe vacancies in the high temperature
phase.%
} after subtracting very small ferromagnetic contribution of 0.003(2)\,$\mu_{B}$/f.u.
due to impurities (probably pure Fe) \cite{Felner_11}. It is the
sum of a paramagnetic part due to magnetic excitation of electrons
near the Fermi surface and a diamagnetic part to which all electrons
contribute. The diamagnetic volume susceptibility is given by the
Langevin equation,
\begin{equation}
\chi_{dia}=-(e^{2}/6Vmc^{2})\sum_{i}Z_{i}\left\langle r^{2}\right\rangle _{i}\label{eq:Langevin}
\end{equation}
The sum is over all the atoms in the unit cell of volume \emph{V},
$\left\langle r^{2}\right\rangle $$_{i}$ is the mean-square radius
of the \emph{i}$^{th}$ atom\textquoteright{}s electron wave function,
and \emph{Z}$_{i}$ denotes atomic number. The diamagnetic contribution
to the magnetization calculated using Eq. \ref{eq:Langevin} equals
to \textminus{}0.014 $\mu_{B}$/ f.u., the paramagnetic part of the
magnetization therefore equals to 0.101(2)-(-0.014)\,=\,0.115(2)\,$\mu_{B}$/
f.u. which is indicated by the diamond symbol in Fig. \ref{fig2}
\footnote{The obtained value is slightly larger than the interpolated \textbf{Q
}(0) value of the magnetic form factor. This difference might be due
to the presence of large diffuse scattering at \emph{T} = 600 K or
might be due to over-estimation of the diamagnetic contribution since
it is apparently calculated for free atoms. Therefore, we have not
included the \textbf{Q }(0) value for the form factor refinement and
maximum entropy calculation.%
}.

\subsection{Unpolarized neutron diffraction}

\begin{table*}
\caption{Parameters obtained in least-squares refinements of integrated intensities
measured at \emph{T}\,=\,600 K on D9.}

\begin{ruledtabular} %
\begin{tabular}{ccccccccc}
Atom  & \multicolumn{4}{c}{Position in \emph{I}4/\emph{mmm}} & \textbf{$\beta_{11}=\beta_{22}$} & $\beta_{33}$ & $\beta_{12},$$\beta_{13}$,$\beta_{23}$ & \multicolumn{1}{c}{\emph{n}}\tabularnewline[\doublerulesep]
\noalign{\vskip\doublerulesep}
 & site  & \emph{x}  & \emph{y} & \emph{z} & \multirow{1}{*}{} &  &  & \tabularnewline[\doublerulesep]
\hline
\noalign{\vskip\doublerulesep}
K  & 2\emph{a}  & 0  & 0  & 0 & 0.13(2) & 0.010 (1) & 0 & 0.82(4) \tabularnewline
Fe & 4\emph{d}  & $\frac{1}{2}$ & 0  & $\frac{1}{4}$ & 0.056(2) & 0.0059(3) & 0 & 0.81(1)\tabularnewline
Se & 4\emph{e}  & 0  & 0  & 0.3532(2) & 0.060(3) & 0.0052(3) & 0 & 1 \tabularnewline[\doublerulesep]
\multicolumn{9}{c}{Extinction \emph{g} (rad$^{-1}$) = 2$\pm$2, \emph{R}$_{F^{2}}$,
\emph{R}$_{wF^{2}}$, \emph{R}$_{F}$, $\chi$$^{2}$: 6.8, 7.3, 6.2,
3.0}\tabularnewline
\multicolumn{9}{c}{\emph{a} = 3.945 (2) Å, \emph{c} = 14.163(4) Å}\tabularnewline
\hline
 &  &  &  &  &  &  &  & \tabularnewline
\multicolumn{9}{c}{Definitions: Thermal factor \emph{T}\,(\emph{h k l}) = exp\{-($\beta_{11}h^{2}$+$\beta_{22}k^{2}+\beta_{33}l^{2}+2\beta_{12}hk+2\beta_{13}hl+2\beta_{23}kl$)\}}\tabularnewline
\multicolumn{9}{c}{$R_{F^{2}}=100\frac{\sum_{n}[\left|G_{obs,n}^{2}-\sum_{k}G_{calc,k}^{2}\right|]}{\sum_{n}G_{obs,n}^{2}}$,
$R_{wF^{2}}=100\sqrt{\frac{\sum_{n}w_{n}(G_{obs,n}^{2}-\sum_{k}G_{calc,k}^{2})^{2}}{\sum_{n}w_{n}G_{obs,n}^{4}}}$,
$R_{F}=100\frac{\sum_{n}[\left|G_{obs,n}-\sqrt{\sum_{k}G_{calc,k}^{2}}\right|]}{\sum_{n}G_{obs,n}}$,}\tabularnewline
\multicolumn{9}{c}{Where the index \emph{n} runs over the observations and the index
\emph{k} runs over the reflections contributing to the obsevation
\emph{n}.}\tabularnewline
\multicolumn{9}{c}{\emph{G}$^{2}$ is the square of the structure factor. $w_{n}=1/\sigma_{n}^{2}$,
is the weight where $\sigma_{n}^{2}$ is the variance of \emph{G}$_{obs,n}$.}\tabularnewline
\end{tabular}\end{ruledtabular} \label{structural_parameters}
\end{table*}
In order to characterize the structural and magnetic phase transitions,
we have measured the temperature dependence of the nuclear ($\frac{3}{5}$\,$\frac{1}{5}$\,0)
and magnetic ($\frac{2}{5}$\,$\frac{1}{5}$\,1) superstructure
peaks at a vertical magnetic field of 9 T oriented along the $\left\langle 1\,1\,0\right\rangle $
direction as shown in Fig. \ref{fig1} (b). The rapid increase of
the intensity of the nuclear ($\frac{3}{5}$\,$\frac{1}{5}$\,0)
peak below \emph{T}$_{\textup{S}}$\,=\,580(3) K indicates the structural
phase transition from the high-temperature Fe-vacancy-disordered phase
with \emph{I}4/\emph{mmm} symmetry into the low-temperature Fe-vacancy
ordered phase with \emph{I4/m} symmetry. The transition temperature
is broadly consistent with that determined from magnetic susceptibility
measurement. The intensity of the magnetic peak ($\frac{2}{5}$\,$\frac{1}{5}$\,1)
vanishes above the antiferromagnetic ordering temperature, \emph{T}$_{\textup{N}}$\,=\,553(3)
K, of the Fe moments. All the measurements for the determination of
magnetization distribution were performed at \emph{T} = 600 K which
is well above both the structural and magnetic phase transitions.
Sets of experimental structure factors containing 80 independent reflections
within sin$\theta$/$\lambda$$\leq$ 0.80 Å$^{-1}$ were obtained
from the integrated intensities measured on D9 after averaging the
intensities over equivalent reflections with a weighted \emph{R$_{wF^{2}}$
}%
\footnote{The weighted \emph{R} factor for the equivalent reflections is defined
as : $R_{wF^{2}}=100[\frac{\sum_{n}w_{n}(I_{obs,n}-I_{mean})^{2}}{\sum_{n}w_{n}I_{mean}^{2}}]^{1/2}$.\emph{
}$w_{n}=1/\sigma_{n}^{2}$, is the weight where $\sigma_{n}^{2}$
is the variance of \emph{I.}%
} factor of 4\%. These data were used in least-squares refinements
of the crystal structure using FullProf \cite{Rodrigues} in which
the variable parameters were the \emph{z} coordinate of Se, the anisotropic
temperature factors for the three sites, a single extinction parameter
\emph{g} representing the mosaic spread of the crystal and the site
occupancies of the K and Se. The results are summarized in Table \ref{structural_parameters}.
The small value obtained for \emph{g}, which is less than its estimated
error, shows that any extinction, if present, is very small. The results
are consistent with the neutron powder diffraction results of Bao
\emph{et al. }on a similar chemical composition \cite{Bao_11}.

\subsection{Polarized neutron diffraction studies}

The flipping ratios were measured at an applied magnetic field of
9 T at 600 K at D3. Since the susceptibility of K$_{0.8}$Fe$_{1.6}$Se$_{2}$
is small ($\sim$ 0.1 $\mu_{B}$/f.u.) all the flipping ratios \emph{R}
are close to unity. Therefore, every reflection was measured for more
than one hour to have reasonable counting statistics. The flipping
ratios measured for equivalent reflections and for repeated measurements
of the same reflection were averaged together to give a mean value
of \emph{R}, which was used to calculate the magnetic structure factors
\emph{F}$_{M}(\mathbf{Q})$ using Cambridge Crystallographic Subroutine
Library (CCSL) \cite{CCSL}. Both the observed flipping ratios and
the calculated \emph{F}$_{M}(\mathbf{Q})$ are listed in Table\,\ref{msf}.
The nuclear structure factor, \emph{F}$_{N}(\mathbf{Q})$, was calculated
using the parameters obtained from the integrated intensity measurements
which are given in Table \ref{structural_parameters}.

\begin{figure}
\centering{}\includegraphics[clip,width=0.45\textwidth]{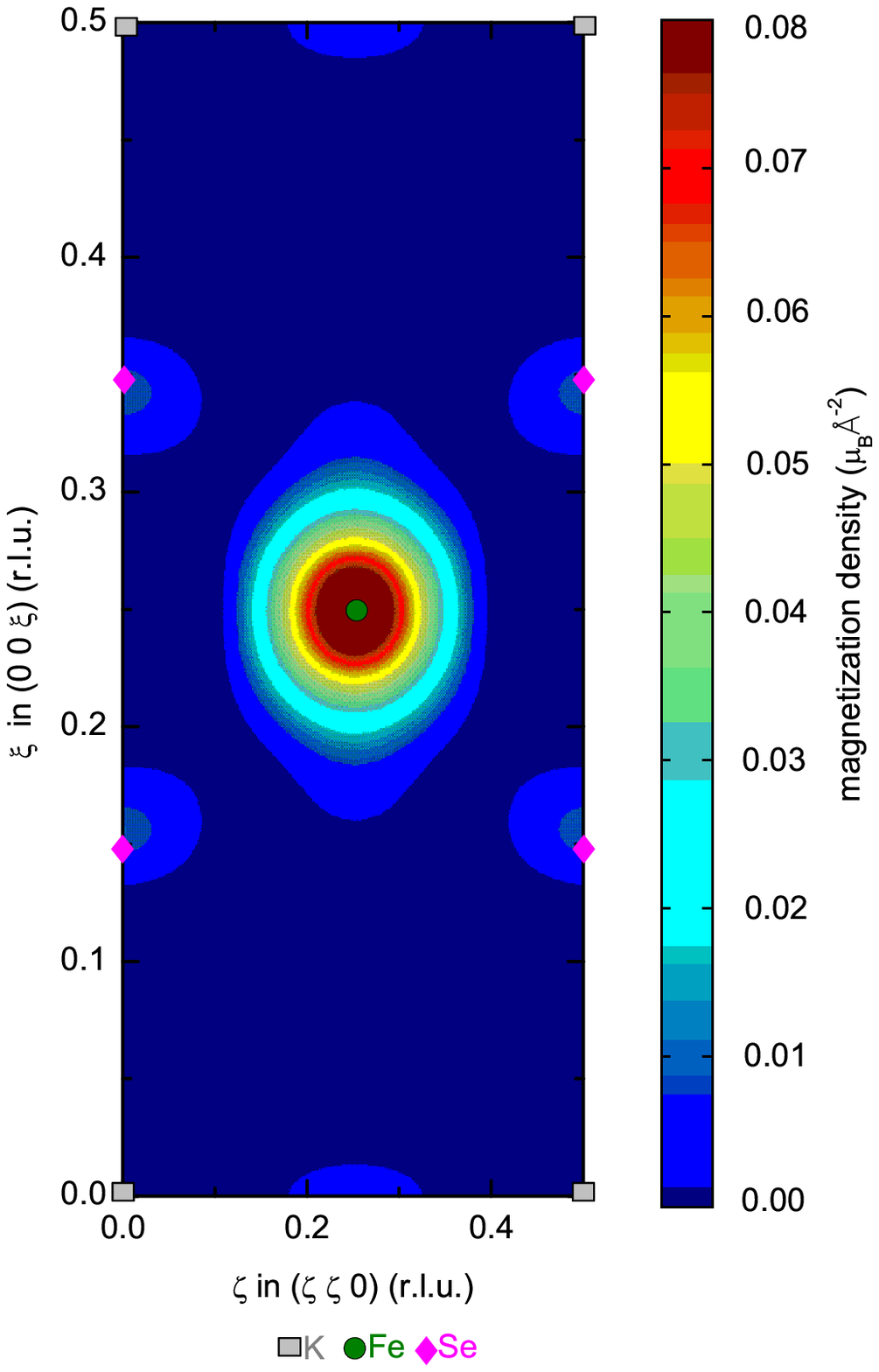}\\
 \caption{\label{fig3} (Color online) Maximum-entropy reconstruction of the
magnetization distribution in tetragonal K$_{0.8}$Fe$_{1.6}$Se$_{2}$
at 600 K projected down to {[}1 1 0{]}.}
\end{figure}

The diamagnetic contribution to the magnetic structure factor in an
applied magnetic field \emph{B} is
\begin{equation}
F_{dia}=\frac{BC}{\left|\mathbf{Q}\right|}\sum_{i}\frac{df_{i}(Q)}{dQ}\exp(i\mathbf{Q}\cdot\mathbf{r}_{i})\label{eq:fdia}
\end{equation}
 where \emph{f$_{i}(Q)$} is the atomic form factor of the \emph{i}$^{th}$
atom and \textbf{r}$_{i}$ its position in the unit cell \cite{Charge_form_factors}.
The constant \emph{C} has the value 1.52$\times$10$^{-5}$\,$\mu_{B}$T$^{-1}$Å$^{2}$
\cite{Statsis_70}. The diamagnetic contributions to the magnetic
structure factors were calculated using Eq. \ref{eq:fdia} and are
given in Table \ref{msf}. The values \emph{F}$_{dia}$ were subtracted
from the magnetic structure factors \emph{F}$_{M}$ to obtain the
paramagnetic structure factors \emph{F}$_{para}$ which are also listed
in Table \ref{msf}. The geometric structure factor of Fe atoms for
the (\emph{h k l}) reflections can be written as:

\begin{equation}
F_{geo}=2\cos(\pi l/2)(e^{i\pi h}+e^{i\pi k})
\end{equation}
which is either +4 or -4 depending on the values of \emph{h, k, l}
assuming full occupancy of the Fe site. In Fig.\,\ref{fig2}, we
show the effective paramagnetic scattering amplitude, obtained by
dividing each \emph{F}$_{para}$ by $F_{geo}\times T(h\, k\, l)$
and multiplied by the number of Fe atoms in the unit cell. It can
be seen that most of the paramagnetic scattering amplitude lies reasonably
close to theoretical spherical form factor $\left\langle j_{0}\right\rangle $
of Fe$^{2+}$ \cite{magnetic_form_factors}. This result signifies
that most of the paramagnetic scattering amplitude is associated with
the Fe$^{2+}$ moments.

\begin{table*}
\caption{Observed and calculated magnetic structure factors for the K$_{0.8}$Fe$_{1.6}$Se$_{2}$
at 600 K and at \emph{B} = 9 T.}

\begin{ruledtabular} %
\begin{tabular}{ccrcrrrrr}
\multicolumn{1}{c}{} &  &  & $\sin$$\theta$/$\lambda$ & (1-\emph{R}) & F$_{M}$ & F$_{dia}$ & F$_{para}$ & F$_{calc}$%
\footnote{using a constrained multipole model with an orbital moment.%
}\tabularnewline
\multicolumn{1}{c}{\emph{h}} & \emph{k}  & \emph{l} & Å$^{-1}$ & $\times10^{3}$ & m$\mu_{B}$/f.u. & m$\mu_{B}$/f.u. & m$\mu_{B}$/f.u. & m$\mu_{B}$/f.u.\tabularnewline
\hline
 0 & 0  & 2  & 0.0706  & 26.8 $\pm$ 0.3  & -92.9$\pm$ 1.2 & 1.13  & -94.0$\pm$ 1.2 & -94.0\tabularnewline
0 & 0  & 4  & 0.1412 & 97.9$\pm$2.2 & 74.9$\pm$ 1.6 & -0.02 & 74.9$\pm$ 1.6 & 77.0\tabularnewline
1 & 1  & 0  & 0.1791 & -101.8$\pm$ 2.9 & -61.1$\pm$ 1.8 & -0.46 & -60.6$\pm$ 1.8 & -69.0\tabularnewline
1 & 1  & 2  & 0.1925 & 28.6$\pm$ 0.5 & 69.3 $\pm$1.1 & -0.42 & 69.8 $\pm$1.1 & 66.0\tabularnewline
1 & 1  & 4  & 0.2281 & 12.5$\pm$ 0.6 & -56.6$\pm$ 2.6 & 0.61  & -57.2$\pm$ 2.6 & -55.0\tabularnewline
2 & 0  & 2  & 0.2630 & 10.3$\pm$ 0.9 & -53.1$\pm$ 4.4 & 0.27 & -53.4$\pm$ 4.4 & -45.0\tabularnewline
1 & 1  & 6  & 0.2774 & 9.8$\pm$ 0.5 & 43.2$\pm$ 2.4 & -0.53  & 43.8$\pm$ 2.4 & 42.0\tabularnewline
0 & 0 & 8  & 0.2824 & 10.2$\pm$ 0.6  & 36.6$\pm$ 2.3 & -0.43 & 37.4$\pm$ 2.3 & 37.0\tabularnewline
2 & 0  & 4  & 0.2900 & 47.5$\pm$ 7.8 & 42.6$\pm$ 7.7 & -0.04 & 42.6$\pm$ 7.7 & 39.0\tabularnewline
1 & 1  & 8  & 0.3345 & 49.4$\pm$ 7.8 & -33.9 $\pm$5.2 & 0.03 & -33.9 $\pm$5.2 & -29.0\tabularnewline
0 & 0  & 10  & 0.3531 & 6.5$\pm$ 0.8 & -22.3$\pm$ 2.6 & 0.28 & -22.6$\pm$ 2.6 & -22.0\tabularnewline
2 & 2  & 0  & 0.3583 & 5.0 $\pm$ 0.8 & 21.6 $\pm$3.3 & -0.32 & 21.9 $\pm$3.3 & 27.0\tabularnewline
2 & 2  & 2  & 0.3651 & 12.0 $\pm$ 2.1 & -27.2 $\pm$4.7 & 0.14 & -27.3 $\pm$4.7 & -26.0\tabularnewline
2 & 0  & 8  & 0.3794 & 5.0$\pm$ 1.1 & 17.6$\pm$ 3.7 & -0.20 & 17.8$\pm$ 3.7 & 22.0\tabularnewline
2 & 2  & 4  & 0.3851 & 45.9$\pm$ 13.6  & 18.7 $\pm$5.4 & -0.01 & 18.7 $\pm$5.4 & 23.0\tabularnewline
3 & 1  & 2  & 0.4067 & 12.8$\pm$ 3.3 & 22.7 $\pm$5.9 & -0.07 & 22.8 $\pm$5.9 & 18.0\tabularnewline
0 & 0 & 12  & 0.4237 & 9.6$\pm$ 2.3 & 14.9 $\pm$3.7 & -0.10 & 15.0 $\pm$3.7 & 11.0\tabularnewline
3 & 1  & 4  & 0.4247 & 3.9 $\pm$ 1.3 & -13.6$\pm$ 4.4 & 0.16 & -13.8$\pm$ 4.4 & -16.0\tabularnewline
2 & 0  & 10 & 0.4345 & 5.1$\pm$ 1.2 & -16.7$\pm$ 3.8 & 0.16 & -16.9$\pm$ 3.8 & -14.0\tabularnewline
3 & 1  & 6 & 0.4531 & 2.6$\pm$ 1.3 & 8.1$\pm$ 4.0 & -0.13 & 8.2$\pm$ 4.0 & 14.0\tabularnewline
2 & 2  & 8  & 0.4562 & 4.6 $\pm$ 1.5 & 9.9 $\pm$3.1 & -0.11 & 10.0 $\pm$3.1 & 14.0\tabularnewline
1 & 1  & 12  & 0.4600 & 8.1 $\pm$ 2.8 & -8.0 $\pm$2.8 & 0.05 & -8.1 $\pm$2.8 & -13.0\tabularnewline
3 & 1  & 8  & 0.4901 & 12.5$\pm$ 21.7 & -7.5 $\pm$13.0 & 0.01 & -7.5 $\pm$13.0 & -10.0\tabularnewline
2 & 0  & 12 & 0.4936 & 7.0$\pm$ 6.5 & 9.9$\pm$ 8.6 & -0.05 & 10.0$\pm$ 8.6 & 8.0\tabularnewline
2 & 2  & 10  & 0.5030 & 3.6 $\pm$ 1.1 & -7.8 $\pm$2.5 & 0.10 & -7.9 $\pm$2.5 & -10.0\tabularnewline
1 & 1  & 14  & 0.5257 & 3.3 $\pm$ 1.2 & 6.3$\pm$ 2.2 & -0.08 & 6.4$\pm$ 2.2 & 5.0\tabularnewline
3 & 3  & 2  & 0.5420 & 6.0 $\pm$ 3.1 & 5.3 $\pm$2.7 & -0.02 & 5.3 $\pm$2.7 & 8.0\tabularnewline
2 & 2  & 12  & 0.5548 & 4.5 $\pm$ 3.1 & 4.3 $\pm$2.9 & -0.03 & 4.3 $\pm$2.9 & 6.0\tabularnewline
3 & 3  & 4 & 0.5556 & 3.2 $\pm$ 1.9 & -6.0$\pm$ 3.6 & 0.06 & -6.0$\pm$ 3.6 & -7.0\tabularnewline
\end{tabular}\end{ruledtabular}

\label{msf}
\end{table*}
To have a model free reconstruction of magnetization density, we have
used maximum entropy method using MEMSYS III subroutine library \cite{Gull}.
This method has been shown to give more reliable results from sparse
and noisy data compared to conventional Fourier analysis \cite{Papoular_90}.
We have used this method to clarify the shape of the distribution.
Figure \ref{fig3} shows maximum entropy reconstruction of the magnetization
distribution projected down to the {[}1 1 0{]} plane using the measured
magnetic structure factors for the {[}H H L{]} type of reflections.
The reconstruction shows clearly that the majority of magnetization
is confined to the region around the iron atoms. However, a very small
magnetization ($\leq0.01$\,$\mu_{B}$Å$^{-2})$ can be seen around
the Se atoms signifying a possible hybridization between the Fe and
Se. The magnetization around the Fe atom is slightly nonspherical
with a shape that appears to extend in the $\left\langle 0\,0\,1\right\rangle $
direction of the projection.

\begin{table}
\caption{Fitting results of the measured form factors with different models
as described in the text and the corresponding agreement factors. }

\begin{ruledtabular} %
\begin{tabular}{ccccc}
Model  & ($\mu_{s}$+$\mu_{l})$$\left\langle j_{0}\right\rangle $ & $\mu_{l}$$\left\langle j_{2}\right\rangle $ & $\chi^{2}$ & \emph{R}$_{w}$\tabularnewline[\doublerulesep]
 & in $\mu_{B}$/Fe & in $\mu_{B}$/Fe  &  & \tabularnewline[\doublerulesep]
\hline
\noalign{\vskip\doublerulesep}
dipole & 0.0320(5) & 0 & 3.3 & 12.6\%\tabularnewline
 & 0.0306(4) & 0.014(4) & 2.0 & 7.9\%\tabularnewline
multipole%
\footnote{Multipole parameters constrained to give only \emph{t}$_{2g}$ type
orbitals.%
} & 0.0319(4) & 0 & 3.4 & 12.5\%\tabularnewline[\doublerulesep]
 & 0.0306(4) & 0.014(4) & 2.2 & 7.7\%\tabularnewline[\doublerulesep]
\hline
\multicolumn{5}{c}{$R_{w}=100\frac{\sum_{n}w_{n}\left|F_{obs,n}^{2}-F_{calc,n}^{2}\right|}{\sum_{n}w_{n}F_{obs,n}^{2}}$.\emph{
}$w_{n}=1/\sigma_{n}^{2}$, is }\tabularnewline[\doublerulesep]
\multicolumn{5}{c}{the weight where $\sigma_{n}^{2}$ is the variance of $F_{obs,n}$}\tabularnewline[\doublerulesep]
\end{tabular}\end{ruledtabular} \label{model}
\end{table}

Further analysis of the measured paramagnetic scattering amplitude
was obtained by fitting the magnetic structure factors to a multipole
model in which they are expressed as

\begin{eqnarray}
F_{M}(\mathbf{Q}) & = & a_{0}\left\langle j_{0}\left|\mathbf{Q}\right|\right\rangle Y(00)+\mu_{L}\left\langle j_{2}\left|\mathbf{Q}\right|\right\rangle \nonumber \\
 &  & +\sum_{l=2,4}\left\langle j_{l}\left|\mathbf{Q}\right|\right\rangle \sum_{m=-l}^{m=l}a_{lm}Y_{\hat{\mathbf{Q}}}(lm\pm)\label{form factor}
\end{eqnarray}
where, $a_{0}=\mu_{s}+\mu_{l}$ is the total magnetic moment of Fe.
$\mu_{s}$ and $\mu_{l}$ are the spin and orbital contributions respectively.
$\left\langle j_{l}\left|\mathbf{Q}\right|\right\rangle $ are the
form factor for a Fe$^{2+}$ and \emph{Y}$_{\hat{\mathbf{Q}}}(lm\pm)$
are the real combinations of spherical harmonic functions written
as,

\begin{equation}
Y_{\hat{\mathbf{Q}}}(lm\pm)=\frac{1}{\sqrt{2}}[Y_{l}^{-m}(\hat{Q)}\pm(-1)^{m}Y_{l}^{m}(\hat{Q)]}
\end{equation}

The point group symmetry of Fe site, $\overline{4}m2,$ limits the
nonzero coefficients \emph{a}$_{lm}$ to \emph{a}$_{20},$ \emph{a}$_{40}$
and \emph{a}$_{44}$. Different models have been considered for fitting
the paramagnetic scattering amplitude in Fig. \ref{fig2}, namely,
(a) a dipole model with only first two terms in the Eq. \ref{form factor},
and (b) a multipole model when all the terms in the Eq. \ref{form factor}
are retained. Both the dipole and multipole models have been considered
with and without the orbital moment. The results of different fitting
models have been summarized in Table \ref{model}. It can be easily
seen that the inclusion of orbital part increases the quality of fit
in both the low and high \textbf{Q} regions of the form factor and
the agreement factors significantly. The obtained ratio of $\frac{\mu_{l}}{\mu_{s}}\approx0.88$
signifies dominant contribution of the orbital moment to the total
paramagnetic scattering amplitude of Fe$^{2+}$.

In a site with fourfold symmetry the \emph{d} orbitals split into
the singlet states: $d_{3z^{2}-r^{2}},$ $d_{x^{2}-y^{2}},$ and $d_{xy}$
and a doublet combination of $d_{xz}$ and $d_{yz}$. The first two
singlet states are derived from the cubic \emph{e}$_{g}$ functions
and the third singlet and the doublet from the \emph{t}$_{2g}$ ones.
The occupancies of these four nondegenerate orbitals can be derived
directly from the coefficients \emph{$a_{lm}$} \cite{Brown_10}.
However, the parameters obtained from the unconstrained fit lead to
unphysical, negative occupancies for the $d_{xy}$ orbital with large
estimated standard deviations for all the orbitals as can be seen
from Table \ref{occupancies}. Band structure calculations as well
as photoemission spectroscopy measurements indicate that the \emph{t}$_{2g}$
orbitals dominate at the Fermi surface \cite{Yi_13,Qian_11,Zhang,Liu_2012}.
Therefore, a constrained fit \cite{Brown_10} in which the ratio between
the \emph{a}$_{lm}$ was fixed to correspond to occupancy of the \emph{t}$_{2g}$-type
orbitals only, gave equally good agreement factor as well as less
standard deviation of the fitted parameters as shown in Table \ref{occupancies}.
The refinement shows that $\sim$\,66\% of the electrons occupy doubly
degenerate \emph{d$_{xz}$/d$_{yz}$} orbitals and $\sim$\,34\%
of those are in the \emph{d$_{xy}$} orbital. The magnetic structure
factors calculated for this constrained multipole model are given
together with the measured values and the diamagnetic corrections
in Table \ref{msf}.

There have been a few reports of magnetization distribution for the
superconducting and non superconducting Fe based compounds \cite{Brown_10,Prokes_11,Lester_11,ratcliff_10,Lee_10}.
Of particular interest are the results of BaFe$_{2}$As$_{2}$ by
Brown \emph{et al}. \cite{Brown_10} who have shown that most of the
magnetization is associated with the Fe atoms and the distribution
is non-spherical with an extension along the $\left\langle 1\,1\,1\right\rangle $
direction. For the superconducting Ba(Fe$_{1-x}$Co$_{x}$)$_{2}$As$_{2}$
samples both Prokeš \emph{et al}. \cite{Prokes_11} and Lester \emph{et
al}. \cite{Lester_11} have concluded that the magnetization is rather
extended along the $\left\langle 1\,1\,0\right\rangle $ direction.
The change in distribution between the doped and undoped samples is
due to the doping induced modifications of the relevant bands near
the Fermi surface as suggested by Lester \emph{et al}. \cite{Lester_11}.
In contrast to all of the investigated Fe-pnictides, K$_{0.8}$Fe$_{1.6}$Se$_{2}$
shows distribution elongated along the $\left\langle 0\,0\,1\right\rangle $
direction. The results of the present experiment show that at least
96\% of the electrons in K$_{0.8}$Fe$_{1.6}$Se$_{2}$, which give
rise to the paramagnetic susceptibility, are localized on the Fe atoms
with a radial distribution similar to that of a Fe$^{2+}$. Their
angular distribution shows that they occupy the \emph{t}$_{2g}$-type
orbitals with a strong preference for the doubly degenerate \emph{xz/yz}
type which is in agreement with the slight elongation observed along
the $\left\langle 0\,0\,1\right\rangle $ direction for the maximum
entropy map in Fig.\,\ref{fig3}. For the BaFe$_{2}$As$_{2}$, the
\emph{xy }($\sim$\,52\%) orbitals are more occupied than the \emph{xz/yz
}($\sim$\,48\%) orbitals which is opposite to the results for the
K$_{0.8}$Fe$_{1.6}$Se$_{2}$ as shown in Table\,\ref{occupancies}.
This difference in occupancy for the \emph{t}$_{2g}$ orbitals and
the corresponding magnetization distribution in different compounds
might be due to the subtle interplay between the crystal field effects
and Hund's rule coupling \cite{Yin_11}.

Most surprising result of the present study is the presence of an
orbital moment to the total paramagnetic moment of K$_{0.8}$Fe$_{1.6}$Se$_{2}$.
Previous studies also hinted existence of orbital moment \cite{Prokes_11,Lester_11}.
For the superconducting Ba(Fe$_{1-x}$Co$_{x}$)$_{2}$As$_{2}$ samples,
Lester \emph{et al}. \cite{Lester_11} have found that at least $\frac{2}{3}$
of the normal state susceptibility does not vanish at the lowest achievable
temperature of 2 K. They have attributed the remaining susceptibility
to the orbital contribution of Van Vleck type or due to the presence
of residual quasiparticle density of states at the Fermi surface.
Our very accurate form factor measurement allows us to quantify the
orbital contribution relative to the spin contribution. The observed
form factor of Fe$^{2+}$ is best fitted using $\sim$\,46\% orbital
and $\sim$\,54\% spin contributions to the total magnetization.
The amount of orbital contribution is unusually high keeping in mind
that the orbital moment is generally quenched in a 3\emph{d}-orbital
system due to the crystal field effects. Nevertheless, similarly large
orbital contribution has been found for the Vanadium \emph{d}-electrons
in classical \emph{s}-wave superconductor V$_{3}$Si using polarized
neutrons \cite{shull}. For the Fe-based superconductors, band structure
calculations also predicted that the orbital contribution is larger
than the spin contribution \cite{Su_11}. Our results provide first
experimental hint of non-negligible orbital contribution to the total
paramagnetic susceptibility of K$_{0.8}$Fe$_{1.6}$Se$_{2}$. Strongly
anisotropic and weakly temperature dependent magnetic susceptibility
observed in the \emph{A}$_{x}$Fe$_{2-y}$Se$_{2}$ (\emph{A\,=\,}K,
Rb, Cs) systems \cite{Lei_11,Liu_11} might be related to the presence
of large orbital contribution since the spin contribution is strongly
temperature dependent.

\begin{table}
\caption{Multipole amplitudes and 3\emph{d }orbital occupancies determined
from the signed magnetic structure factors using CCSL.\emph{ }}

\begin{ruledtabular} %
\begin{tabular}{cccc}
 &  &  & Amplitudes (in $\mu_{B}$/Fe)\tabularnewline[\doublerulesep]
\cline{4-4}
Function & coefficient & all \emph{d}%
\footnote{All multipole parameters allowed by the $\overline{4}m2$ point group
symmetry.%
} & \emph{t}$_{2g}$ only%
\footnote{Multipole parameters constrained to give only \emph{t}$_{2g}$ type
orbitals.%
}\tabularnewline[\doublerulesep]
\hline
\noalign{\vskip\doublerulesep}
Y(00) & \emph{a}$_{0}$ & 0.0306(5) & 0.0306(4)\tabularnewline
Y(20) & \emph{a}$_{20}$ & 0.04(10) & -0.03(9)\tabularnewline
Y(40) & \emph{a}$_{40}$ & -0.20(40) & -0.39(6)\tabularnewline[\doublerulesep]
Y(44+) & \emph{a}$_{44}$ & 0.48(58) & -0.37(7)\tabularnewline[\doublerulesep]
 & $\mu_{L}$ & 0.017(4) & 0.014(4)\tabularnewline[\doublerulesep]
$\chi^{2}$ &  & 2.2 & 2.2\tabularnewline[\doublerulesep]
 & \multicolumn{3}{c}{Occupancies (\%)}\tabularnewline[\doublerulesep]
\cline{2-4}
orbital & \multicolumn{2}{c}{all \emph{d$^{a}$}} & \emph{t$_{2g}$ }only$^{b}$\tabularnewline[\doublerulesep]
\hline
3\emph{z$^{2}$-r$^{2}$} & \multicolumn{2}{c}{12(24)} & 0\tabularnewline[\doublerulesep]
\emph{x$^{2}-y^{2}$} & \multicolumn{2}{c}{40(35)} & 0\tabularnewline[\doublerulesep]
\emph{xy} & \multicolumn{2}{c}{-6(35)} & 34(7)\tabularnewline[\doublerulesep]
\emph{xz, yz} & \multicolumn{2}{c}{54(30)} & 66(7)\tabularnewline[\doublerulesep]
\end{tabular}\end{ruledtabular} \label{occupancies}
\end{table}

\section{Conclusion}

In summary, we have determined magnetization distribution in K$_{0.8}$Fe$_{1.6}$Se$_{2}$
using polarized neutron diffraction. Magnetic structure factors derived
from the polarization dependence of the intensities of the Bragg reflections
were used to make a maximum-entropy reconstruction of the distribution
projected on the {[}1\,1\,0{]} plane. The reconstruction shows clearly
that the magnetization is confined to the region around the iron atoms.
A very small magnetization around the Se atoms hints towards a possible
hybridization between the Fe and Se. The distribution of magnetization
around the Fe atom is slightly nonspherical with a shape which is
extended in the $\left\langle 0\,0\,1\right\rangle $ direction in
the projection. These results show that the electrons which give rise
to the paramagnetic susceptibility are confined to the Fe atoms and
their distribution suggests that they occupy 3\emph{d} \emph{t}$_{2g}$-type
orbitals with 66\% in those of \emph{xz/yz} symmetry. Orbital moment
contributes significantly to the total paramagnetic moment of Fe and
might be responsible for the anisotropic properties of the Fe-based
superconductors.

\bibliographystyle{apsrev} \bibliographystyle{apsrev}
\begin{acknowledgments}
S. N. likes to acknowledge M. Angst for helpful discussions and M.
Vial and A. John for technical assistance.\bibliographystyle{apsrev}
\bibliography{SmFeAsO}
\end{acknowledgments}

\end{document}